# Structural and Magnetic Properties of Barium Hexaferrite Nanoplatelets

Daniel Zabek*[a,b], Joseph Veryard[b], Yusra Ahmed[b], Arjen van den Berg[c], Joseph Askey[c], and Sam Ladak[c]

[a]School of Engineering, University of Southampton, SO17 1BJ, Southampton, UK
[b]School of Engineering, Cardiff University, CF24 3AA, Cardiff, UK.
[c]School of Physics and Astronomy, Cardiff University, CF24 3AA, Cardiff, UK.

*email: D.A.Zabek@soton.ac.uk

**Abstract:** Combining unique geometric and magnetic anisotropy in barium hexaferrites has the potential to enhance the performance of advanced technological applications, such as data storage, spintronics, and medicine. Here, we report the synthesis and deposition of barium hexaferrite nanoplatelets, followed by comprehensive structural and magnetic microscopies. The topographic, physical, and morphological properties of individual nanoplatelets, clusters, and aggregates are analyzed using atomic force microscopy (AFM) and electron microscopy, while magnetic properties are analysed using magnetic force microscopy (MFM). Relevant size distributions and nanoparticle configurations for magnetic interactions have been experimentally identified and numerically analyzed. For dense nanoparticle arrangements, direct exchange coupling dominates with parallel magnetisation configuration in overlapping particles. In contrast, separated particles exhibit anti-parallel coupling. Detailed micro-magnetic modelling elucidates the length scales over which these two interaction regimes are dominant, paving the way for macroscopic two-dimensional (2D) and three-dimensional (3D) structures with tuned magnetic ordering.

1. Introduction

Nanoscale magnetic materials offer great potential to address key challenges in catalysis and medicine with recent breakthoughs in neuromodulation and immunotherapy[1-7]. With their high surface-to-volume ratio, nanomagnetic materials are also considered the building block for larger three-dimensional (3D) magnetic structures with applications in spintronics and nanomagnetics[8-10]. Additionally, these materials exhibit a variety of geometries, including spheres, rods or wires, platelets, and core-shell structures, all with unique static and dynamic magnetic properties[11]. Unlike spherical nanoparticles, the anisotropic geometry of rods and platelets is of particular interest because their aspect ratio influences magnetic properties, which can be precisely controlled for the desired application[12]. In this field, ferrites represent a highly versatile class of materials, enabling numerous compounds with specific performance characteristics. A particularly promising ferrimagnetic ferrite system is $BaFe_{12}O_{19}$ (Barium hexaferrite, or BHF), which has high saturation magnetization (72 Am² kg⁻¹), high coercivity, high Curie temperature (450 °C), as well as excellent chemical stability[13]. In particular, scandium-substituted BHF magnetic nanoplatelets (MNPs) are of significance due to their strong geometric and magnetocrystalline anisotropy. With BHF platelets of one or two unit cells stacked (~2 - 5 nm thick) perpendicular to their basal plane, such systems can be approximated as a two-dimensional (2D) materials. Furthermore, their high saturation magnetisation and strong magnetocrystalline anisotropy oriented along the crystallographic *c*-axis facilitates the development of next generation recording technologies and RF absorbers[14,15]. Anisotropic BHF MNPs dispersed in isotropic liquids also show particularly striking magnetic domain formations as liquid crystals, enabling novel magneto-optical principles for bioimaging[16-18]. When utilizing BHF MNPs for 2D or 3D magnetic applications, it is crucial to first achieve control over the individual nanoparticles, their aggregates, and their interactions with the desired substrate[19]. In this letter, we report on the synthesis of ferrimagnetic BHF nanoplatelets and the systematic nanoscale characterization using electron microscopy and surface probe scanning techniques, followed by a comparison of their magnetic characteristics with micromagnetic modelling. Topographic analysis and visualization of magnetic domain switching using atomic and magnetic force microscopy, respectively, are performed on nanomagnetic BHF particles and aggregates. The results obtained here contribute to numerous research and design activities in the field of solid and liquid magnetic structures using anisotropic MNPs.

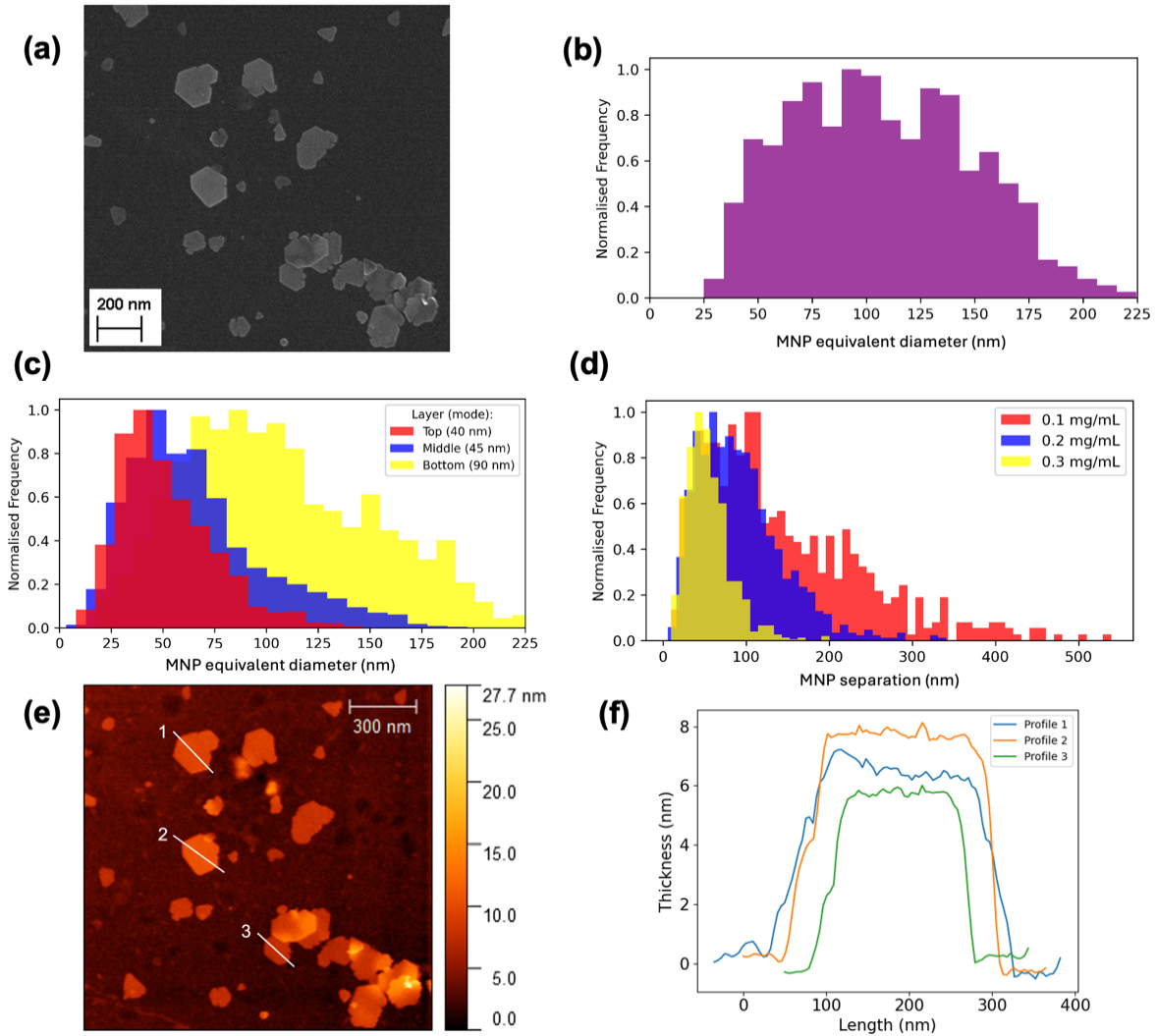

**Figure 1:** (a) SEM scan of Sc-BHF MNPs. Size distributions of (b) bulk and (c) centrifuged MNPs with (d) interparticle separation distance distribution. (e) AFM image of BHP particles with (f) AFM topography profiles shown for selected MNPs. Blue, orange, and green represent profiles 1, 2, and 3 respectively in (e).

2. Materials and Methods

A: Nanoparticle synthesis

Sc-substituted BHF nanoplatelets were synthesised hydrothermally at 240 °C, following a previously reported procedure, and functionalized using dodecylbenzenesulfonic acid (DBSA) surfactant[20]. The BHF nanoplatelets were dispersed in 1-butanol (99.8%) and sonicated for two hours, creating a stable ferrofluid suspension.

B: Nanoparticle deposition

The nanoplatelets were deposited via drop-casting onto a carbon film on a copper TEM grid (Agar Scientific) and heated on a hot plate at 60°C for 20 minutes. The 1-butanol carrier fluid fully evaporated, resulting in a grid with flat, dry nanoparticles coated with a 1 nm layer of DBSA[21]. **Figure 1(a)** shows a Scanning Electron Microscopy (SEM) image of Sc-BHF MNPs. Based on scans of 400 individual nanoparticles, the equivalent diameter (i.e. a disk diameter with the same corner-to-corned distance) of the synthesized hexagonal-shaped MNPs was measured yielding the size distribution shown in **Figure 1(b)**, with average equivalent median diameter of 90 nm. To vary the concentration and size distribution of the deposition, the ferrofluid suspension was centrifuged at 3,372 relative centrifugal force (rcf) for 105 minutes, resulting in the separation of MNPs. This separation was feasible due to the narrow thickness distribution of Sc-BHF MNPs[15,22]. Three distinct phases were formed at the top, middle, and bottom of the centrifuge vial (see Supplementary Figure 1), with corresponding modal diameters of 40 nm (top), 45 nm (middle), and 90 nm (bottom). The size distributions of these phases are shown in **Figure 1(c)**, illustrating a shift towards larger equivalent diameters as the centrifugal force increases[23]. The effect of suspension concentration on the drop-cast particles was studied by analyzing the inter-particle separation distances, defined as the shortest distance between adjacent particles of each nanoparticle's nearest neighbor on the TEM grid, as determined from SEM scans. **Figure 1(d)** shows modal separation distances of 50, 70, and 110 nm for corresponding suspension concentrations of 0.3, 0.2, and 0.1 mg/mL, respectively. Increasing the particle concentration above 0.3 mg/mL leads to significant particle aggregation, as shown in SEM scans (see Supplementary Figure 2). MNP substrate surface coverage can thus be tuned up to 0.3 mg/mL without inducing particle aggregation. By adjusting the suspension concentration, the population density and surface coverage of MNPs on the substrate can be controlled, reducing empty space and thereby increasing interactions between particles. To identify the implications of particle size and separation upon magnetic properties, individual particles and aggregates were measured using atomic and magnetic force microscopy. Magnetic force microscopy (MFM) imaging is performed by first capturing the topography of the surface using atomic force microscopy (AFM). The probe is then lifted by a user-defined height to measure phase shifts in the cantilever oscillations induced by long-range dipolar interactions between the surface and a magnetised tip, allowing for magnetic sampling[24]. This method produces both topographic and magnetic images of the sample, enabling correlation of magnetic properties with structural features. AFM scans of the surface profile area selected in Figure 1(a) were obtained using a commercial Bruker Dimension Icon with a ScanAsyst-Air probe, while MFM imaging was performed with a commercial magnetic probe (MESP-V2, Bruker) using a Co-Cr tip. According to the manufacturers data sheet, the tip has a magnetic moment of $1\times10^{-16}$ $Am^2$, a coercivity of 40 mT, and a nominal radius of curvature of 35 nm. Prior to MFM imaging, the tip was subjected to a 0.5 T magnetic field along the long axis of tip, using a neodymium permanent bar magnet. **Figure 1(e)** shows the structural features of the MNPs deposited on the grid, as observed with AFM. **Figure 1(f)**, shows the average discrete thicknesses of the measured DBSA coated particles to be 5.3, 6.4, and 7.6 nm (±0.02 nm), aligning with the RSRS stacking sequence of alternating structural R-type and S-type building blocks of a M-type ferrite molecular unit[14,25]. The AFM data also determined approximately 70% of nanoparticles, possessed a 5.3 nm thickness and the remaining 20% and 10% were 6.4 and 7.6 nm, respectively. This agrees with the expected uniform thicknesses for Sc-BHF fabricated via hydrothermal synthesis[22]. The AFM surface profile scan further confirms the relatively smooth topography of the coated MNPs, with a mean average roughness value Ra of 0.4 nm, suggesting a continuous unit cell structure. Based on the proposed synthesis method, the particle diameter has a significant effect on the saturation magnetization of Sc-BHF MNPs powders[15].

C: Micromagnetic model

The magnetic properties of individual MNPs remain difficult to extract for the median size distribution determined in Figure 1(b). For this reason, finite-difference micromagnetic simulations are used to study single particles and pairs of idealized hexagonal MNPs focusing on the preferred magnetization configurations between two MNPs at remanence. All simulations are conducted using Ubermag[26,27] with OOMMF[28] as a computational backend. The total energy $E$ of the system considered is the integral of the exchange $w_{ex}$, demagnetization $w_d$, and the uniaxial magnetocrystalline anisotropy $w_{mc}$ energy densities over the total volume $V$ of the MNP(s):

$$E = \int (w_{ex} + w_d + w_{mc})\, dV \qquad (1)$$

The various interactions manifest in an effective field $\mathbf{H}_{eff}$ acting on the magnetization of the system:

$$\mathbf{H}_{eff} = -\frac{1}{\mu_0 M_s}\frac{\partial E}{\partial \mathbf{m}} \qquad (2)$$

The dynamics of the system are driven through precession and damping of the magnetization, $\mathbf{m} = \mathbf{M}/M_s$, about the effective field, described by the Landau-Lifshitz-Gilbert equation:

$$\frac{\partial \mathbf{m}}{\partial t} = \frac{-\gamma_0}{1+\alpha^2}\left(\mathbf{m}\times\mathbf{H}_{eff} + \alpha\,\mathbf{m}\times(\mathbf{m}\times\mathbf{H}_{eff})\right) \qquad (3)$$

with gyromagnetic ratio $\gamma_0$ = 2.211×10⁵ mA⁻¹s⁻¹ and dimensionless Gilbert damping $\alpha$ = 0.01[29]. Material parameters are based on BHF spherical MNPs[30] with saturation magnetization $M_s$ = 2.75×10⁵ Am⁻¹, exchange constant $A_{ex}$ = 2.00×10⁻¹¹ Jm⁻¹, and uniaxial magnetocrystalline anisotropy constant $K_u$ = 9.60×10⁴ Jm⁻³ with uniaxial anisotropy easy axis along the c-axis, perpendicular to the basal plane of the MNP. For a MNP perfectly flat on the substrate, the c-axis and cartesian z-axis are equivalent. The mesh size in the basal plane (i.e the xy-directions) is chosen to be 1 nm in order to resolve true lateral shape which is an order of magnitude smaller than the exchange length $l_{ex} = \sqrt{2A_{ex}/\mu_0 M_s^2}$ = 20.5 nm and the exchange-anisotropy length (domain wall width parameter) $l_{an} = \sqrt{A_{ex}/K}$ = 14.4 nm. In the easy axis direction (i.e the *z*-direction) the mesh size is chosen to be 5 nm to alleviate computational time. Idealised pairs of MNPs with a thickness of *t* = 5 nm and corresponding experimental diameter of *d* = 90 nm is simulated for a varying range of edge-to-edge MNP separations *s* by appropriate mapping of the saturation magnetization only in regions bounded by the edges of the MNPs, and zero everywhere else in the simulation space.

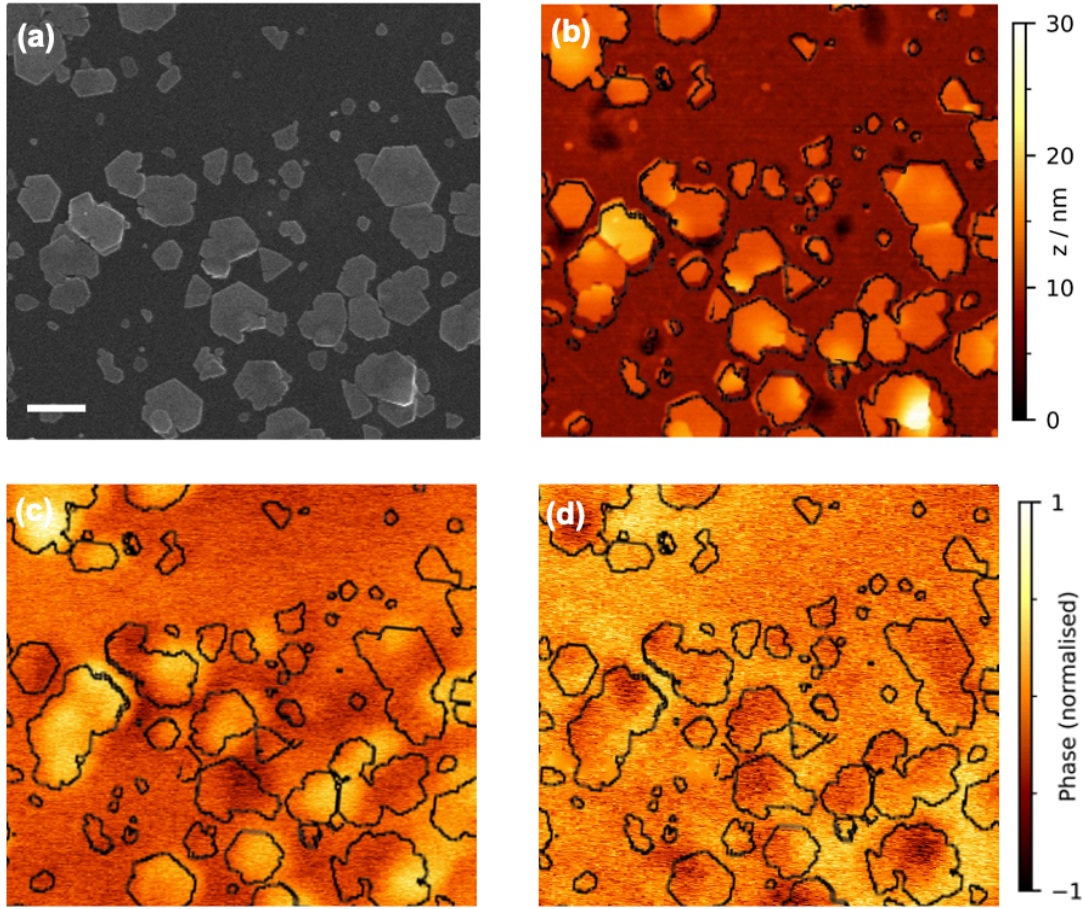

**Figure 2:** Sc-BHF MNPs characterized via SEM (a), AFM (b), MFM (c) and reverse tip MFM following probe magnetization inversion (d). The MNP perimeters are highlighted to improve readability with a 100 nm scale bar in (a) also used for figures (b-d).

3. Results and Discussion

A: Magnetic Force Microscopy

In **Figure 2(a)**, an SEM scan of Sc-BHF MNPs is shown, fabricated following the same protocol as in Figure 1(a). MNP perimeters in Figure 2(b-d) are outlined to emphasise correllations between SEM, AFM, MFM scans. We note that smaller particles are not resolved in the atomic force micrograph shown in **Figure 2(b)** due to resolution limits determined by the MESP-V2 tip radius of 35 nm. MFM data is shown in **Figures 2(c-d)** and after magnetsing the tip along two opposite directions. Here, long-range dipolar interactions between the magnetic nanoparticles and the magnetised scanning probe microscope probe introduce positive and negative phases shifts in the cantilever oscillations which are visualised by bright and dark contrast (respectively) in **Figure 2(c)**. Reversing the tip magnetisation inverts contrast attributable to magnetostatic interactions as shown in **Figure 2(d)** whilst leaving artefacts due to electrostatic interactions unchanged, providing confirmation of the magnetic origin of the observed contrast. Uniform contrast in small single isolated MNPs indicate a single-domain state without internal magnetic structure. However, those with larger diameters or with overlap, appear to have some internal structure. Finally, some particles do not seem to have inverted contrast likely due to either the particle being non-magnetic and the contrast arising solely from changing electrostatic interactions attributable to the particle topography, or the particle magnetisation switching due to the tip stray field. Here, a perpendicularly aligned MESP-V2 tip with a minimum lift height of 44 nm exhibits a magnetic field of 235 mT in *z*-direction. In contrast, coercivity values for BHF are reported to be between 100 mT for powders to 650 mT for particles, which is lower than the tip magnetic field[31-33]. The uniform magnetic contrast in selected particles also suggests that these MNPs are magnetized in and out of plane, indicating uniaxial magnetocrystalline anisotropy with an easy axis perpendicular to their basal plane. Given a lateral resolution approximately equal to the MFM pass lift-height, MNPs smaller than 50 nm are not resolvable in the MFM measurement. It can be predicted that the superparamagnetic limit of BHF must be below the tip resolution and is approximated for idealised spherical MNPs to be 11.2 nm[34]. Although the study in Figure 1(c-d) on centrifugation and ferrofluid concentration provides some control over single BHF MNP coverage, the true effect of MNP aggregation on magnetic properties and device applicability is not fully understood[35].

B: Magnetic switching

To investigate the nature of any coupling between adacent nanoparticles, the sample was exposed to and external magnetic field of +2 T perpenedicular to the substrate direction (i.e. Helmholtz coil geometry) and then scanned using MFM. Subsequently, the sample was exposed to an external magnetic field of -2 T and MFM scanned again. Here, the applied external fields are greater than the coercive fields required for magnetic switching.

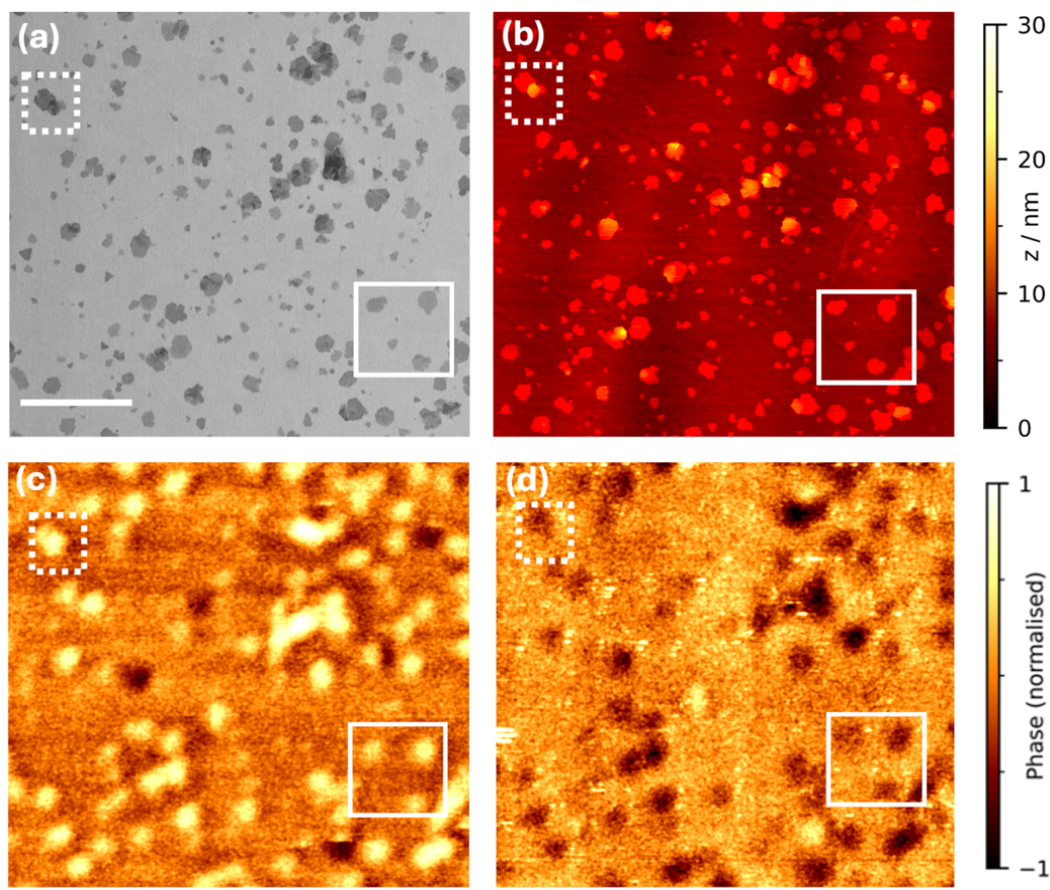

**Figure 3:** (a) STEM of region with dispersed BHF nanoparticles. (b) Corresponding AFM image. (c) MFM image acquired after application of a 2 T magnetic field in the positive *z*-direction. (d) MFM image acquired after application of a 2 T magnetic field in the negative *z*-direction. The small dotted outline shows two particles that exhibit antiferromagnetic coupling whilst the solid large outline shows separated single domain particles with parallel magnetisation. Scale bar represents 1 µm.

**Figure 3(a)** shows a scanning transmission electron microscopy (STEM) micrograph of the relevant sample area, with separated MNPs highlighted in the large boxed inset and MNPs in close proximity or stacked configuration with other MNPs highlighted in the small dashed inset. The corresponding AFM scan is shown in **Figure 3(b)**. **Figure 3(c-d)**, shows that not all MNPs remain magnetised in the field direction following the +2 T and -2 T sequence, which can be attributed to damaged MNPs, impurities, thermal fluctuations, cracks, or incomplete synthesis on the particle surface, leading to a non-collinear spin configuration[34,36]. However, separated MNPs in the large solid-line boxed inset have parallel magnetisation, while some overlapped MNPs in the small dashed inset exhibit antiparallel magnetisation. On the other hand, stacked MNPs exhibit a significantly stronger signal (i.e. phase shift) compared to separated MNPs, which suggests parallel coupling in MNP aggregates.

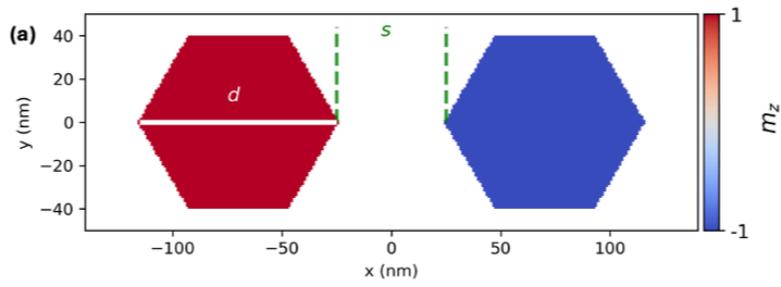
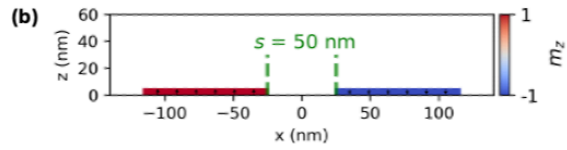
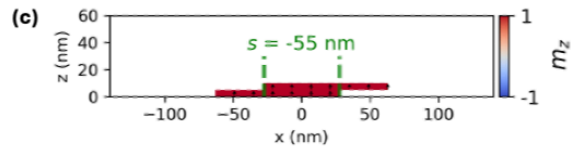
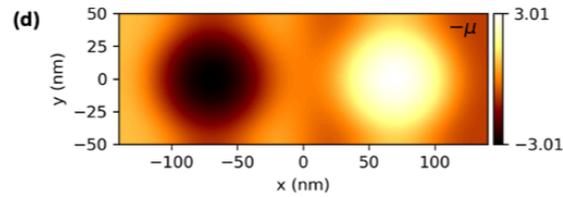
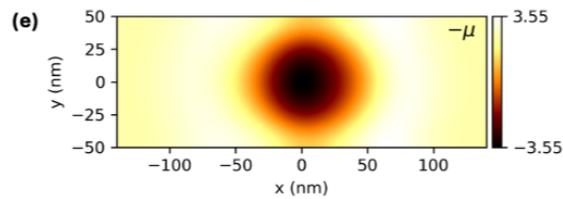
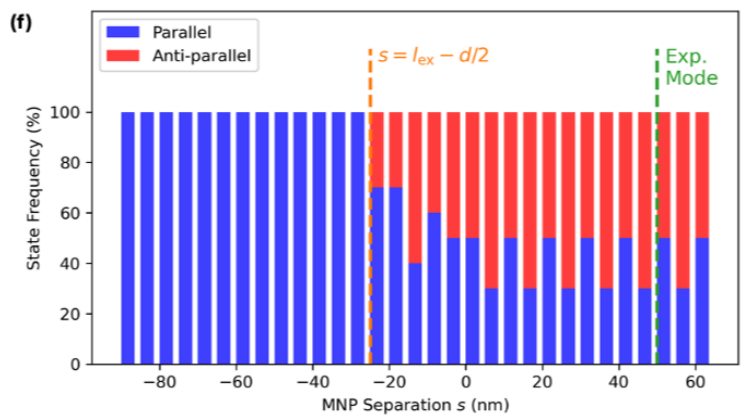
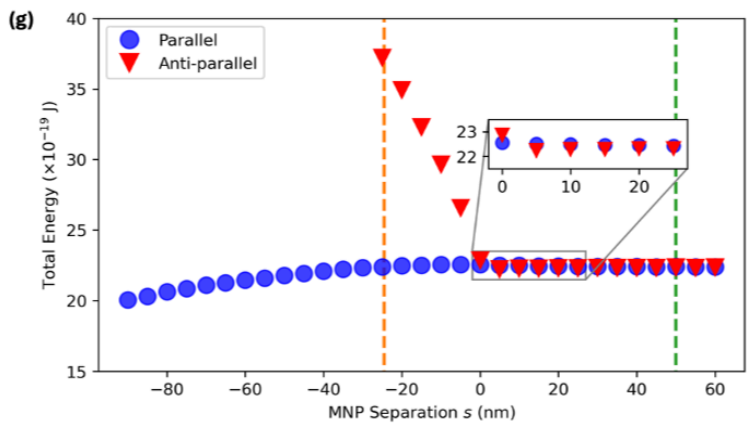

**Figure 4:** Micromagnetic simulations of MNPs. (a) Schematic of the relaxed anti-parallel magnetisation configuration of two MNPs with outer diameter $d$ and edge-to-edge separation $s$. (b-c) Schematics of the MNP side-profiles for MNPs with $d$ = 90 nm and $s$ = 50 nm and for $s$ = - 55 nm where MNPs are offset in $z$ by thickness $t$. (d, e) Simulated MFM phase shifts for $s$ = 50 nm for both (d) positive, i.e +$z$, and negative, i.e -$z$, tip magnetic dipole moments and (e) MFM phase shifts for $s$ = - 55 nm. (f) The corresponding state frequency and energetics of two adjacent nanoparticles, as a function of edge-to-edge separation $s$. (g) Illustrates the magnetisation alignment configuration statistics for $d$ = 90 nm. Blue corresponds to parallel magnetisation alignment which is dominant for $s < l_{ex} - d/2$ (orange dashed lines). Red corresponds to anti-parallel magnetisation alignment, which appears for $s \geq l_{ex} - d/2$ for $d/2$ = 90 nm.

C: Micromagnetic simulations

**Figure 4(a)** shows a schematic of MNPs, which have been subject to a micromagnetic study. For values where the particle seperation $s$ < 0, one of the MNPs is offset in the $z$-direction by thickness $t$ to emulate a stacked configuration, as shown in **Figure 4(b-c)**. The initial magnetization distributions in the MNPs are randomized, and the system is allowed to relax to an minimum energy state. This process is repeated with different random initializations for a total of 10 relaxations for each value of $s$, providing a statistical outlook on the preferred relaxed state as a function of separation. The expected MFM phase shift is simulated in **Figure 4(d-e)** based on experimental parameters with a lift height $z_{lift}$ = 44 nm, magnetic dipole moment $\mu$ = ±1×10$^{-16}$ Am$^{-2}$ spring constant $k$ = 3 Nm$^{-1}$, quality factor $Q$ = 192[37] and a Guassian convolution of the tip with the sample defined by full-width half maximum FWHM = 35 nm equal to the nominal tip radius of curvature given by the manufacturer. Comparing the experimental data shown in Figure 3(c-d) with the micromagnetic simulations shown in Figure 4(d-e) for two isolated idealized MNPs with $d$ = 90 nm and $s$ = +50 nm and $s$ = -55 nm, it can be noted that the simulated MFM in Figure 4 compares well with the experimental MFM in Figures 3(c-d). The surrounding regions of the MNPs also give rise to a non-zero phase shift due to the stray fields extending from the basal plane and the edges of the MNPs. The phase shift for the surrounding region of the MNPs is generally opposite to that of the MNP and weaker in magnitude. This is replicated in the experimental MFM, where dark (bright) contrast is surrounded by weaker bright (dark) contrast in Figures 3(c-d), respectively. This arises due to the sign, magnitude, and second derivatives of the $z$-component of the stray fields $H_z$ for separated and overlapping MNPs, which are plotted in Supplementary Figure 3. Here, the second derivatives at the edges alternate between positive and negative values, and again within the centers of the MNPs, giving rise to a weaker and opposite phase shift for separated MNPs. For overlapping MNPs, the magnitude of the phase shift is larger for the overlapped regions compared to the separated MNPs, again well replicated in the experimental data, where there are clear regions of overlapping MNPs (e.g. the small dashed inset in Figure 3) exhibiting larger phase shift magnitudes compared to regions of well-separated MNPs. From the experimental MFM data in Figure 3(c-d), we can see that pairs and aggregates of MNPs can exhibit either a parallel magnetization configuration, illustrated by bright-bright (dark-dark) contrast, or an anti-parallel magnetization configuration, shown by dark-bright (bright-dark) contrast. Micromagnetic simulations of idealized pairs of hexagonal MNPs were performed to better understand the preferred configurations as a function of the MNP edge-to-edge separation $s$. The statistics for the nominal experimental MNP diameter $d$ = 90 nm are shown in the non-ideal shapes of the MNPs in the experimental systems have not been considered in the simulations, nor has the field-driven history of the MNPs, which would influence the exact energy minimum state of the MNPs. The field-driven metastability of the parallel and anti-parallel configurations, non-idealized MNP shapes, and the addition of more than two MNPs within the system are subjects for further investigation. Finally, it is vital to optimise the macroscopic properties of the deposition for future studies and applications, as aggregation can potentially retard macroscopic magnetic properties at the device level. In **Figure 4(f)**, parallel configurations are the only observed state for $s < l_{ex} - d/2$, i.e for overlapping MNPs, as shown by the orange dashed line. This can be straightforwardly understood as the result of dominant exchange and uniaxial anisotropy interactions favoring parallel magnetization. For $s > l_{ex} - d/2$, anti-parallel alignment configurations are now observed at first in small percentages of the total relaxation sample set then becoming dominant at $s$ = -15 nm. For values $s$ > 0 nm , the dominant configuration alternates between parallel and anti-parallel, indicating that the energy differences between the two states are small; therefore, relaxation into either state is equally likely, i.e., the parallel and anti-parallel configurations become metastable. It is also notable that the preferred configurations alternate based on whether $s$ is odd or even, suggesting that the preferred state may be influenced by the discretization of the simulation space. **Figure 4(g)** illustrates the energies of the two configurations as a function of $s$, where the lowest energy configuration is the parallel alignment for $s$ < 0 nm, and the anti-parallel alignment for $s$ > 0 nm, as can be shown by the inset plots of the energies up to $s$ = 25 nm. The percentage difference in energy is on the order of 1%, and for the largest values of $s$ the percentage differences are on the order of 0.1%. These results indicate that we could expect to see equal numbers of parallel and anti-parallel configurations in the experimentally realized systems; however, there is a clear preference for parallel alignments of the MNPs, especially when exposed to and external magnetic field in Figure 3(c-d). Whilst our AFM and MFM studies demonstrate regimes where coupling matches simulations, interactions are also likely to be sensitive to local variations in topography, chemical disorder and strain which surface probe microscopy cannot resolve. In addition, future challenges remain for the use of BHF in balancing short-range electrostatic forces (e.g., ionization and van der Waals forces) in suspension with long-range magnetostatic interactions during deposition. While the proposed deposition technique has been optimized for flat substrates, the identified magnetic coupling interactions of MNPs can be extended into 3D space. This advancement supports the development of self-assembled and functionalized MNPs in solid arrays and liquid programmable matter.

4. Conclusions

The geometrical and magnetic anisotropy of scandium-substituted barium hexaferrite, along with the characterization of nanoplatelets and aggregates, remains of fundamental interest for the development of macroscopic structures. Here, we established control over the deposition process and discussed for the first time the effects on magnetic domain coupling using electron microscopy (SEM/STEM) and surface probe microscopy (AFM/MFM) techniques. The corresponding single domains were accurately determined for nanoparticles with an equivalent diameter greater than 50 nm. When utilizing scandium-substituted barium hexaferrite nanoplatelets, the magnetic easy axis is perpendicular to the basal plane, which can be directly observed via MFM. For small nanoplatelet separation distances, anti-parallel coupling is preferred, while for stacked aggregates, only parallel coupling has been observed. Given the current limited control over device fabrication, future work will include precise deposition and lithography to enable multi-particle functionality and three-dimensional nanomagnetics in solid and liquid systems.

Acknowledgement

This research received funding from the Royal Academy of Engineering (RAEng) under the Research Fellowship Program Number RF\201819\18\202. SL acknowledges funding from the Leverhulme Trust (RPG-2021-139) and the Engineering and Physical Sciences Research Council (EP/R009147/1 and EP/X012735/1)

Conflict of interest

There are no conflicts to declare.

References

[1]     Mohammed, L., Gomaa, H.G., Ragab, D. and Zhu, J., 2017. Magnetic nanoparticles for environmental and biomedical applications: A review. *Particuology,* 30, pp.1-14.

[2]     Zhang, Q., Yang, X. and Guan, J., 2019. Applications of magnetic nanomaterials in heterogeneous catalysis. *ACS Applied Nano Materials*, *2*(8), pp.4681-4697.

[3]     Stueber, D.D., Villanova, J., Aponte, I., Xiao, Z. and Colvin, V.L., 2021. Magnetic

nanoparticles in biology and medicine: past, present, and future trends. *Pharmaceutics*, *13*(7), p.943.

[4] Wu, K., Su, D., Liu, J., Saha, R. and Wang, J.P., 2019. Magnetic nanoparticles in nanomedicine: a review of recent advances. *Nanotechnology*, *30*(50), p.502003.

[5] Hescham, S.A., Chiang, P.H., Gregurec, D., Moon, J., Christiansen, M.G., Jahanshahi, A., Liu, H., Rosenfeld, D., Pralle, A., Anikeeva, P. and Temel, Y., 2021. Magnetothermal nanoparticle technology alleviates parkinsonian-like symptoms in mice. *Nature communications*, *12*(1), p.5569.

[6] Choi, S.H., Shin, J., Park, C., Lee, J.U., Lee, J., Ambo, Y., Shin, W., Yu, R., Kim, J.Y., Lah, J.D. and Shin, D., 2024. In vivo magnetogenetics for cell-type-specific targeting and modulation of brain circuits. *Nature Nanotechnology*, *19*(9), pp.1333-1343.

[7] Li, Y., Wu, W., Liu, Q., Wu, Q., Ren, P., Xi, X., Liu, H., Zhao, J., Zhang, W., Wang, Z. and Lv, Y., 2024. Specific surface-modified iron oxide nanoparticles trigger complement-dependent innate and adaptive antileukaemia immunity. *Nature Communications*, *15*(1), p.10400.

[8] Bhowmik, D., You, L. and Salahuddin, S., 2014. Spin Hall effect clocking of nanomagnetic logic without a magnetic field. *Nature nanotechnology*, *9*(1), pp.59-63.

[9] Mathur, N., Stolt, M.J. and Jin, S., 2019. Magnetic skyrmions in nanostructures of non-centrosymmetric materials. *APL Materials*, *7*(12).

[10] Fernández-Pacheco, A., Streubel, R., Fruchart, O., Hertel, R., Fischer, P. and Cowburn, R.P., 2017. Three-dimensional nanomagnetism. *Nature communications*, *8*(1), pp.1-14.

[11] Roca, A.G., Gutiérrez, L., Gavilán, H., Brollo, M.E.F., Veintemillas-Verdaguer, S. and del Puerto Morales, M., 2019. Design strategies for shape-controlled magnetic iron oxide nanoparticles. *Advanced drug delivery reviews*, *138*, pp.68-104.

[12] Lisjak, D. and Mertelj, A., 2018. Anisotropic magnetic nanoparticles: A review of their properties, syntheses and potential applications. *Progress in Materials Science*, *95*, pp.286-328.

[13] Went, J.J., Rathenau, G.W., Gorter, E.W. and van Oosterhout, G.V., 1952. Hexagonal iron-oxide compounds as permanent-magnet materials. *Physical Review*, *86*(3), p.424.

[14] Pullar, R.C., 2012. Hexagonal ferrites: a review of the synthesis, properties and applications of hexaferrite ceramics. *Progress in Materials Science*, *57*(7), pp.1191-1334.

[15] Lisjak, D., Bukovec, M. and Zupan, K., 2016. Suppression of the exaggerated growth of barium ferrite nanoparticles from solution using a partial substitution of Sc 3+ for Fe 3+. *Journal of Nanoparticle Research*, *18*, pp.1-11.

[16] Shuai, M., Klittnick, A., Shen, Y., Smith, G.P., Tuchband, M.R., Zhu, C., Petschek, R.G., Mertelj, A., Lisjak, D., Čopič, M. and Maclennan, J.E., 2016. Spontaneous liquid crystal

and ferromagnetic ordering of colloidal magnetic nanoplates. *Nature communications*, *7*(1), p.10394.

[17] Gregorin, Ž., Sebastián, N., Osterman, N., Boštjančič, P.H., Lisjak, D. and Mertelj, A., 2022. Dynamics of domain formation in a ferromagnetic fluid. *Journal of Molecular Liquids*, *366*, p.120308.

[18] Hu, J., Gorsak, T., Rodríguez, E.M., Calle, D., Muñoz-Ortiz, T., Jaque, D., Fernández, N., Cussó, L., Rivero, F., Torres, R.A. and Sole, J.G., *Magnetic Nanoplatelets for High Contrast Cardiovascular Imaging by Magnetically Modulated Optical Coherence Tomography. ChemPhotoChem 2019, 3 (7), 529–539*.

[19] Pinilla-Cienfuegos, E., Kumar, S., Mañas-Valero, S., Canet-Ferrer, J., Catala, L., Mallah, T., Forment-Aliaga, A. and Coronado, E., 2015. Imaging the Magnetic Reversal of Isolated and Organized Molecular-Based Nanoparticles using Magnetic Force Microscopy. *Particle & Particle Systems Characterization*, *32*(6), pp.693-700.

[20] Ahmed, Y., Paul, A., Boštjančič, P.H., Mertelj, A., Lisjak, D. and Zabek, D., 2023. Synthesis of barium hexaferrite nano-platelets for ethylene glycol ferrofluids. *Journal of Materials Chemistry C*, *11*(45), pp.16066-16073.

[21] Hribar Boštjančič, P., Tomšič, M., Jamnik, A., Lisjak, D. and Mertelj, A., 2019. Electrostatic interactions between barium hexaferrite nanoplatelets in alcohol suspensions. *The Journal of Physical Chemistry C*, *123*(37), pp.23272-23279.

[22] Makovec, D., Gyergyek, S., Goršak, T., Belec, B. and Lisjak, D., 2019. Evolution of the microstructure during the early stages of sintering barium hexaferrite nanoplatelets. *Journal of the European Ceramic Society*, *39*(15), pp.4831-4841.

[23] Boštjančič, P.H., Gregorin, Ž., Sebastián, N., Osterman, N., Lisjak, D. and Mertelj, A., 2022. Isotropic to nematic transition in alcohol ferrofluids of barium hexaferrite nanoplatelets. *Journal of Molecular Liquids*, *348*, p.118038.

[24] Kazakova, O., Puttock, R., Barton, C., Corte-León, H., Jaafar, M., Neu, V. and Asenjo, A., 2019. Frontiers of magnetic force microscopy. *Journal of applied Physics*, *125*(6).

[25] Makovec, D., Komelj, M., Dražić, G., Belec, B., Goršak, T., Gyergyek, S. and Lisjak, D., 2019. Incorporation of Sc into the structure of barium-hexaferrite nanoplatelets and its extraordinary finite-size effect on the magnetic properties. *Acta Materialia*, *172*, pp.84-91.

[26] Beg, M., Lang, M. and Fangohr, H., 2021. Ubermag: Toward more effective micromagnetic workflows. *IEEE Transactions on Magnetics*, *58*(2), pp.1-5.

[27] Beg, M., Pepper, R.A. and Fangohr, H., 2017. User interfaces for computational science: A domain specific language for OOMMF embedded in Python. *AIP Advances*, *7*(5).

[28] Donahue, M.J. and Porter, D.G., 1999. *OOMMF User's Guide: Version 1.0* [online].

[29] Li, Z.W., Yang, Z.H., Kong, L.B. and Zhang, Y.J., 2013. High-frequency magnetic properties at K and Ka bands for barium-ferrite/silicone composites. *Journal of magnetism and*

*magnetic materials*, *325*, pp.82-86.

[30] Vázquez Bernárdez, M.J., Vukadinovic, N., Lefevre, C. and Stoeffler, D., 2024. Tuning Dynamic Susceptibility in Barium Hexaferrite Core–Shell Nanoparticles through Size-Dependent Resonance Modes. *ACS Applied Electronic Materials*, *6*(5), pp.3274-3282.

[31] Lisjak, D., Arčon, I., Poberžnik, M., Herrero-Saboya, G., Tufani, A., Mavrič, A., Valant, M., Boštjančič, P.H., Mertelj, A., Makovec, D. and Martin-Samos, L., 2023. Saturation magnetisation as an indicator of the disintegration of barium hexaferrite nanoplatelets during the surface functionalisation. *Scientific reports*, *13*(1), p.1092.

[32] Manglam, M.K., Kumari, S., Mallick, J. and Kar, M., 2021. Crystal structure and magnetic properties study on barium hexaferrite of different average crystallite size. *Applied Physics A*, *127*, pp.1-12.

[33] Kurniawan, C., Widodo, A.T. and Djuhana, D., 2020, July. The diameter effect on the magnetization switching time of sphere-shaped ferromagnets using micromagnetic approach. In *IOP Conference Series: Materials Science and Engineering* (Vol. 902, No. 1, p. 012060). IOP Publishing.

[34] Drofenik, M., Ban, I., Makovec, D., Žnidaršič, A., Jagličić, Z., Hanžel, D. and Lisjak, D., 2011. The hydrothermal synthesis of super-paramagnetic barium hexaferrite particles. *Materials Chemistry and Physics*, *127*(3), pp.415-419.

[35] Gutiérrez, L., De la Cueva, L., Moros, M., Mazarío, E., De Bernardo, S., De la Fuente, J.M., Morales, M.P. and Salas, G., 2019. Aggregation effects on the magnetic properties of iron oxide colloids. *Nanotechnology*, *30*(11), p.112001.

[36] Coey, J.M.D., 1971. Noncollinear spin arrangement in ultrafine ferrimagnetic crystallites. *Physical Review Letters*, *27*(17), p.1140.

[37] Corte-León, H., Neu, V., Manzin, A., Barton, C., Tang, Y., Gerken, M., Klapetek, P., Schumacher, H.W. and Kazakova, O., 2020. Comparison and validation of different magnetic force microscopy calibration schemes. *Small*, *16*(11), p.1906144.

Structural and Magnetic Properties of Barium Hexaferrite Nanoplatelets

Zabek*[a,b], D., Veryard[b], J., Ahmed[b], Y., van den Berg[c], A., Askey[c], J., and Ladak[c], S.

[a]School of Engineering, University of Southampton, SO17 1BJ, Southampton, UK
[b]School of Engineering, Cardiff University, CF24 3AA, Cardiff, UK.
[c]School of Physics and Astronomy, Cardiff University, CF24 3AA, Cardiff, UK.

*email: D.A.Zabek@soton.ac.uk

Following centrifugation, nanoparticles from each region were found to have, confirming the particle size distribution broadens with inceasing centrifugal forces down the vial.

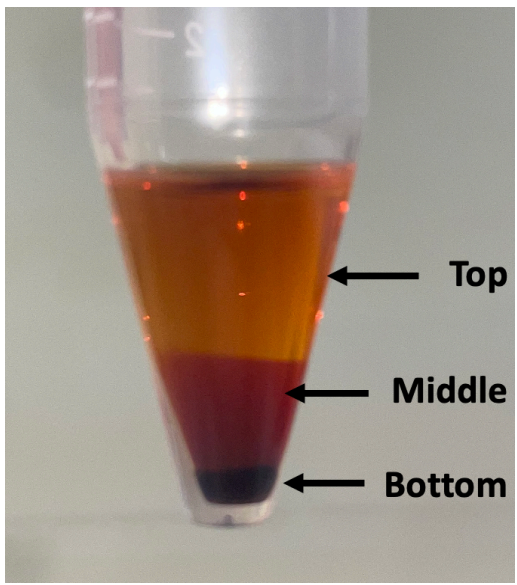

**Supplementary Figure 1:** Three distinct ferrofluid phases formed in a tube after centrifugation.

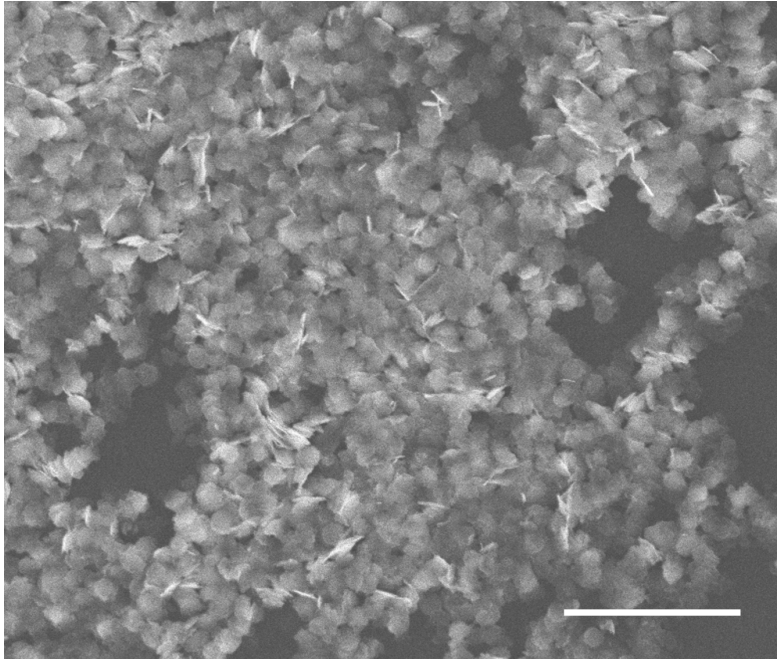

**Supplementary Figure 2**: SEM images showing the aggregation of Sc-BHF MNPs that occurs when deposited at above 1 mg/mL concentration. The scale bare is 1 µm.

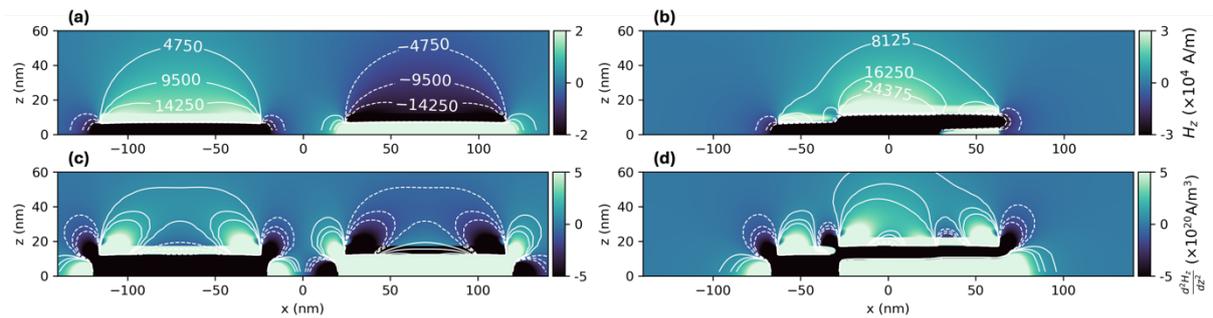

**Supplementary Figure 3**: Micromagnetic simulations of the MNP stray fields as shown in **Figure 2**. (a, c) $H_z$ and $d^2H_z/dz^2$ for separated MNPs with $s$ = 50 nm. (b,d) $H_z$ and $d^2H_z/dz^2$ for overlapping MNPs with $s$ = - 55 nm.